\documentclass[11pt]{article}

\usepackage[margin=1.15in]{geometry}
\usepackage{amsmath,amssymb,amsthm}
\usepackage{natbib}
\usepackage{hyperref}
\usepackage{xcolor}
\usepackage{microtype}
\usepackage{enumitem}
\usepackage{pgfplots}
\pgfplotsset{compat=1.16}

\hypersetup{
    colorlinks=true,
    linkcolor=blue!60!black,
    citecolor=blue!60!black,
    urlcolor=blue!60!black
}

\title{\bfseries Why Constants Matter in Distribution Testing:\\
From Uniformity to Calibration}

\author{Alon Kipnis}

\date{\today}

\newcommand{\ECE}{\mathrm{ECE}}
\newcommand{\Pp}{\mathbb P}

\newcommand{\calA}{\mathcal A}

\newcommand{\Unif}{\mathrm{Unif}}
\newcommand{\PhiG}{\Phi}
\newcommand{\Multi}{\mathrm{Multinomial}}

\begin{document}

\maketitle

\begin{abstract}
Distribution goodness-of-fit testing has developed a powerful rate-level theory: we often know how the required sample size scales with the alphabet size, the separation from the null, and the target error probability. Uniformity testing is the canonical example. One can distinguish the uniform distribution on \(N\) categories from alternatives at total-variation distance at least \(\epsilon\) with far fewer than \(N\) samples, and the optimal scaling is now well understood.

But rate-level theory leaves an important question unresolved: among several tests with the same sample-complexity order, which one actually gives the best risk or power? This is a constant-level question. It is especially relevant in modern applications where distribution testing is used not merely as an asymptotic abstraction, but as a practical design tool.

This note argues that sharp constants in distribution testing play a role analogous to Fisher information in parametric estimation and Pinsker's constant in nonparametric estimation. First, they distinguish between tests that are all rate-optimal but not equally powerful. Second, they reveal the effective signal-to-noise ratio governing the testing problem. Third, they can guide tuning-parameter choices in downstream applications. We illustrate this perspective through large-alphabet uniformity testing and then explain why the same logic matters for choosing the number of bins in calibration testing.

\end{abstract}

\section{Distribution testing beyond rates}

A central lesson from property testing and modern nonparametric statistics is that testing can be much easier than estimation. If \(p\) is an unknown distribution on \(N\) categories, estimating \(p\) in total variation requires order \(N\) samples. By contrast, testing whether \(p\) is uniform can often be done with order \(\sqrt N\) samples, up to the dependence on the separation parameter \(\epsilon\).

The emphasis on constants is familiar from other parts of statistics. In regular parametric estimation, the \(n^{-1/2}\) rate is only the first-order message; the Fisher information determines the sharp asymptotic variance and therefore distinguishes between experiments and estimators with the same rate. Similarly, in nonparametric estimation, minimax rates describe how risk scales with sample size, but Pinsker's constant gives the sharp asymptotic benchmark for the leading minimax risk over smoothness classes; see, for example, \citet{nussbaum1999minimax}. These constants are not cosmetic refinements: they are the quantities that turn asymptotic theory into quantitative risk predictions. The same principle applies in distribution testing. Once the detection rate is known, the next question is the sharp constant governing the limiting risk.

The testing-versus-estimation phenomenon was made especially clear by \citet{paninski2008coincidence}, who showed that in the sparse regime the number of repeated observations, or collisions, contains enough information to detect nonuniformity even when most categories are unobserved.

This rate-level perspective has been extremely successful. It tells us when testing is possible and when it is impossible. However, it does not fully determine the statistical performance of a concrete test. Two tests may both require
\[
    n \asymp \frac{\sqrt N}{\epsilon^2}
\]
samples and yet have noticeably different error probabilities at the same \(n,N,\epsilon\). In other words, sample-complexity rates can identify the right scale while still hiding the most practically relevant comparison.

A sharper question is therefore:
\[
    \textit{What is the exact asymptotic risk, including the leading constant?}
\]
This question is not only a refinement of theory. It changes the way we compare tests.

\section{A simple Gaussian analogy}

The importance of constants is easiest to see in the Gaussian testing problem
\[
    H_0: X \sim N(0,1),
    \qquad
    H_1: X \sim N(u,1).
\]
Here it is natural to decide between $H_0$ and $H_1$ based on whether the test statistic $X$ exceeds a threshold $t$. The risk, denoted $R(t)$, is the sum of the Type-I and Type-II errors:
\[
    R(t) = \Pp_{H_0}(X > t) + \Pp_{H_1}(X \leq t) = (1-\PhiG(t)) + \PhiG(t-u). 
\]
The threshold that minimizes the risk is $t = u/2$, so the minimal risk is
\[
    R(u/2) = (1-\PhiG(u/2)) + \PhiG(-u/2) = 2\PhiG(-u/2).
\]
Thus, the value of the signal-to-noise ratio $u$ directly determines the risk. For example,
\[
    u=2
    \quad\Rightarrow\quad
    2\PhiG(-1) \approx 0.317,
\]
whereas
\[
    u=4
    \quad\Rightarrow\quad
    2\PhiG(-2) \approx 0.046.
\]
From a rate perspective, both are just fixed nonzero signal-to-noise ratios. But from a risk perspective, they correspond to very different testing performance.

Sharp-constant distribution testing seeks analogous formulas for high-dimensional discrete testing problems. The goal is not only to say that a problem becomes solvable when a certain normalized separation diverges, but to identify the precise signal-to-noise ratio that determines the limiting risk.

\section{Uniformity testing as the model problem}

Uniformity goodness-of-fit testing is the cleanest setting in which we ask these questions. The data $(V_1,\ldots,V_n)$ are i.i.d.\ from a distribution $\mathbf{q} = (q_1,\ldots,q_N)$ on $N$ categories, and the null hypothesis is
\[
    H_0 \,:\, q_i = \frac{1}{N},
    \qquad i=1,\ldots,N.
\]
Under the alternative, \({\mathbf q}\) is separated from uniformity in some metric, for example total variation, \(\ell_2\), or more generally \(\ell_p\).

Many test statistics have been proposed for this problem. The most common are the collision test, the chi-squared test, and the empirical total variation test. At the rate level, several of these procedures can be optimal or nearly optimal in appropriate regimes. But they are not equivalent at the constant level. \citet{gupta2022sharp} make this point explicitly in their work on sharp constants for uniformity testing via the Huber statistic: theoretical rate analyses can fail to rank the empirical performance of common uniformity testers, and constant factors explain part of this discrepancy.

This suggests that the right asymptotic object is not only the detection boundary, but the limiting risk curve.

\section{Sharp minimax risk}

A sharp minimax formulation asks for
\[
    R^*(\epsilon,n,N)
    =
    \inf_{\psi}
    \left[
        \Pp_0(\psi=1)
        +
        \sup_{{\mathbf q}\in \calA_\epsilon}
        \Pp_{\mathbf q}(\psi=0)
    \right],
\]
where \(\calA_\epsilon\) is the alternative class and \(\psi\) ranges over all tests. One may also invert the minimax-risk curve to obtain two familiar quantities:
the minimax separation radius and the sample complexity. For a target total
risk \(\gamma\in(0,1)\), define the \(\gamma\)-minimax separation radius at
sample size \(n\) by
\[
\epsilon_\gamma^*(n,N)
=
\inf\left\{
\epsilon>0:
R^*(\epsilon,n,N)\leq \gamma
\right\}.
\]
Similarly, for a fixed separation level \(\epsilon\), define the
\(\gamma\)-sample complexity by
\[
n_\gamma^*(\epsilon,N)
=
\inf\left\{
n\in\mathbb N:
R^*(\epsilon,n,N)\leq \gamma
\right\}.
\]

In large-alphabet uniformity testing against \(\ell_p\)-separated alternatives, \citet{kipnis2026minimax} characterizes this minimax risk at the constant level. 

In the remainder of this note, we specialize to $p=1$, both for simplicity and to match the common $\ell_1$ formulation used in much of the distribution-testing literature. Throughout, $\epsilon$ denotes $\ell_1$ distance from uniformity, $\|\mathbf q-\mathbf u\|_1=\sum_i|q_i-1/N|$, which is twice the total-variation distance.

In the Poissonized model, in the intermediate regime where the risk converges to a nontrivial constant, the limiting risk takes the Gaussian form
\[
    R^*(\epsilon,n,N)
    =
    2\PhiG(-u/2)+o(1),
\]
where \(u\) is the effective signal-to-noise ratio.

With the notation and normalization of \citet{kipnis2026minimax}, this signal-to-noise ratio may be written in the form
\[
    u = u_{\epsilon,n,N} =
    \sqrt{2N}\,
    \sinh\left(
        \frac{n\epsilon^2}{2N}
    \right).
\]
In the small-signal regime, using \(\sinh(x)\sim x\), this becomes
\[
    u_{\epsilon,n,N}
    \sim
    \frac{n\epsilon^2}{\sqrt{2N}}.
\]
Thus the high-dimensional goodness-of-fit problem behaves, at the level of minimax risk, like a one-dimensional Gaussian testing problem. 

The analysis also identifies the asymptotically minimax test. The optimal statistic is a linear functional of the occupancy histogram,
\[
    Z_m = \#\{\text{categories appearing exactly $m$ times in the data}\},
\]
for $m=1,\ldots,n$. In the very sparse regime, the optimal statistic is driven by collisions $Z_2$. In denser regimes, it behaves more like a chi-squared statistic. This provides a constant-level explanation of why collision and chi-squared tests can each be natural, but in different sampling regimes.

\section{What constants reveal that rates hide}

Sharp constants add three pieces of information that rates alone do not provide.

First, they distinguish between tests that have the same detection boundary. If two tests both succeed when
\[
    \frac{n\epsilon^2}{\sqrt N}\to\infty,
\]
rate theory treats them as equivalent. Sharp-risk theory can show that one has a substantially smaller limiting error probability than the other.

Second, constants identify the correct signal-to-noise ratio. In the large-alphabet uniformity problem, the relevant quantity is the effective signal-to-noise ratio \(u_{\epsilon,n,N}\). Once \(u_{\epsilon,n,N}\) is known, a risk target can be translated directly into a sample-size requirement.

Third, constants make asymptotic theory operational. Suppose we want total minimax risk at most \(\gamma\). The Gaussian formula gives the requirement
\[
    2\PhiG(-u/2)\leq \gamma,
\]
or equivalently
\[
    u\geq 2\PhiG^{-1}(1-\gamma/2).
\]
For a level-\(\alpha\) test with Type-II error at most \(\beta\), the analogous Gaussian approximation gives
\[
    u\geq z_{1-\alpha}+z_{1-\beta},
\]
where $z_{1-\alpha}$ is the $1-\alpha$ quantile of the standard normal distribution (i.e., $\PhiG(z_t)=t$).

\section{From uniformity testing to calibration testing}

Calibration assessment in machine learning provides a natural setting in which
sharp constants become operational. Consider a predictive model that outputs a
conditional distribution function \(\hat F(y\mid x)\). Given a test observation
\((X,Y)\), define the probability integral transform (PIT)
\[
    V=\hat F(Y\mid X).
\]
Under perfect distributional calibration, the PIT is uniform:
\[
    H_0:\quad V\sim \Unif(0,1).
\]
Thus testing calibration of a continuous predictive model can be reduced to a
uniformity testing problem for PIT values.

Let \(Q\) denote the distribution function of \(V\), and suppose it has density
\(q\). The Expected Calibration Error (ECE) is a natural population-level calibration distance:
\[
    \ECE(Q)
    := \int_0^1 |q(v)-1|\,dv.
\]
For \(p=1\), this is the area between the PIT density and the uniform density.

In practice, however, calibration is rarely tested directly in the continuum.
The PIT values are binned. If \(N\) equal-width bins are used, with edges
\(0=t_0<t_1<\cdots<t_N=1\), then the bin counts
\[
    O_j = \#\{i: V_i\in [t_{j-1},t_j]\},\qquad j=1,\ldots,N,
\]
are compared to the uniform baseline \(n/N\). Writing
\[
    q_j = Q(t_j)-Q(t_{j-1}),
    \qquad
    \mathbf q=(q_1,\ldots,q_N),
\]
the binned calibration problem becomes the multinomial goodness-of-fit problem
\[
    (O_1,\ldots,O_N)\sim \Multi(n,\mathbf q),
    \qquad
    H_0^{(N)}:\mathbf q=(1/N,\ldots,1/N).
\]
The corresponding discretized calibration error is
\[
    \ECE^{(N)}(\mathbf q) = \sum_{j=1}^N |q_j-1/N|.
\]

This formulation is developed in \citet{kipnis2026calibrating}, which uses
sharp minimax uniformity testing to derive optimal binning and minimax
calibration tests for continuous predictive models.

\section{Binning creates two competing errors}

The number of bins \(N\) is not merely a plotting choice. It is a statistical
resolution parameter. Choosing \(N\) too small creates discretization bias:
localized or oscillatory departures from calibration may cancel inside wide
bins. Choosing \(N\) too large creates statistical noise: the \(n\) test samples
are spread too thinly across bins, and the goodness-of-fit test loses power.

The smoothing effect of binning is visible from the identity
\[
    q_j-\frac1N
    =
    \int_{t_{j-1}}^{t_j} (q(v)-1)\,dv.
\]
By the triangle inequality, we have
\[
    \ECE^{(N)}(\mathbf q)
    =
    \sum_{j=1}^N
    \left|
        \int_{t_{j-1}}^{t_j} (q(v)-1)\,dv
    \right|
    \leq
    \sum_{j=1}^N
    \int_{t_{j-1}}^{t_j} |q(v)-1| \,dv
    = \int_0^1 |q(v)-1|\,dv = \ECE(Q).
\]
Thus,
\[
    \ECE^{(N)}(\mathbf q)\leq \ECE(Q).
\]
Binning can only reduce the apparent calibration error. If \(q(v)-1\) oscillates
within a bin, positive and negative deviations may cancel, making a miscalibrated
predictive model appear nearly calibrated at that binning resolution.

This is the first side of the trade-off. The second side is statistical. For a
fixed binned separation level \(\epsilon\), the relevant signal-to-noise ratio
for minimax calibration testing is, in the small-signal regime,
\[
    u_{\epsilon,n,N}
    \sim
    \frac{n\epsilon^2}
    {\sqrt{2N}}.
\]
Thus, once \(N\) is large enough to resolve the miscalibration, increasing \(N\)
further reduces the signal-to-noise ratio by spreading the same sample size over
more bins.

\section{Sharp constants give an optimal binning rule}

The sharp minimax theory gives a direct way to translate a desired testing
guarantee into a maximum number of bins. For the binned calibration problem,
consider the minimax risk
\[
    R^*
    =
    \inf_{\psi}
    \sup_{\mathbf q:\,\ECE^{(N)}(\mathbf q)\geq \epsilon}
    \left[
        \Pp_0(\psi=1)
        +
        \Pp_{\mathbf q}(\psi=0)
    \right].
\]
In the asymptotic regime where
\[
    u_{\epsilon,n,N}\to u^*\in(0,\infty),
\]
the minimax risk satisfies
\[
    R^*
    =
    2\PhiG(-u^*/2)+o(1).
\]
Moreover, this risk is attained by the statistic
\[
    T_N
    =
    \sqrt{\frac{N}{2n^2}}
    \sum_{j=1}^N
    \left[
        (O_j-n/N)^2-O_j
    \right].
\]
This statistic is closely related to chi-squared, but the subtraction of
\(O_j\) stabilizes the variance in sparse large-\(N\) regimes. 
In particular, the variance of $T_N$ remains constant even if $n$ is random with a Poisson distribution, so $T_N$ is robust to sample-size fluctuations.

Given a target total risk \(R\), the condition
\[
    2\PhiG(-u_{\epsilon,n,N}/2)\leq R
\]
is equivalent to
\[
    u_{\epsilon,n,N}
    \geq
    2\PhiG^{-1}(1-R/2).
\]
Solving this inequality gives the maximum number of bins
\[
    N_{\max}
    =
    \left\lfloor
    \frac{n^2\epsilon^4}
    {8[\PhiG^{-1}(1-R/2)]^2}
    \right\rfloor.
\]
Equivalently, if one wants a level-\(\alpha\) test with worst-case Type-II error
at most \(\beta\), the Gaussian approximation gives the condition
\[
    u_{\epsilon,n,N}
    \geq
    z_{1-\alpha}+z_{1-\beta},
\]
and hence
\[
    N_{\max}
    =
    \left\lfloor
    \left(
        \frac{n^2\epsilon^4}
        {2(z_{1-\alpha}+z_{1-\beta})^2}
    \right)
    \right\rfloor.
\]

This formula makes precise the design principle:
\[
    \textit{use the finest binning resolution that still preserves the desired testing power.}
\]
Using fewer bins may hide continuous miscalibration through bin averaging. Using
more than \(N_{\max}\) may resolve the miscalibration visually but make it
statistically undetectable.

\section{An oscillatory calibration example}

The binning trade-off is especially clear for oscillatory miscalibration. Suppose
a miscalibrated predictive CDF perturbs the true conditional CDF in probability
space according to
\[
    \hat F(y\mid x)
    =
    F(y\mid x)
    +
    \frac{a}{2\pi k}
    \sin\!\bigl(2\pi k F(y\mid x)\bigr),
\]
where \(k\) is the number of oscillation cycles and \(a\in(0,1)\) is the
amplitude. This is a valid CDF perturbation and corresponds to a
miscalibrated quantile map. The PIT density is
\begin{equation}
    \label{eq:oscillatory-density}
    q(v) = 1+a\cos(2\pi k v), \qquad v\in[0,1],
\end{equation}
with CDF $Q(v)=v+\tfrac{a}{2\pi k}\sin(2\pi k v)$.
This density oscillates around the uniform density. Its continuous
\(\ell_1\)-calibration error is
\[
    \epsilon_\infty
    := \int_0^1 |q(v)-1|\,dv = \frac{2a}{\pi}.
\]

However, after binning into \(N\) equal-width bins, the apparent calibration
error is attenuated:
\[
    \ECE^{(N)}(\mathbf q)
    \approx
    \epsilon_\infty
    \left|\mathrm{sinc}(k/N)\right|,
    \qquad
    \mathrm{sinc}(x)=\frac{\sin(\pi x)}{\pi x}.
\]
Thus, when \(N\) is too small relative to the oscillation frequency \(k\), the
miscalibration cancels inside bins and the binned ECE is close to zero. When
\(N\) is very large, the oscillation is resolved, but the expected count per bin
\(n/N\) becomes small and sampling noise dominates.

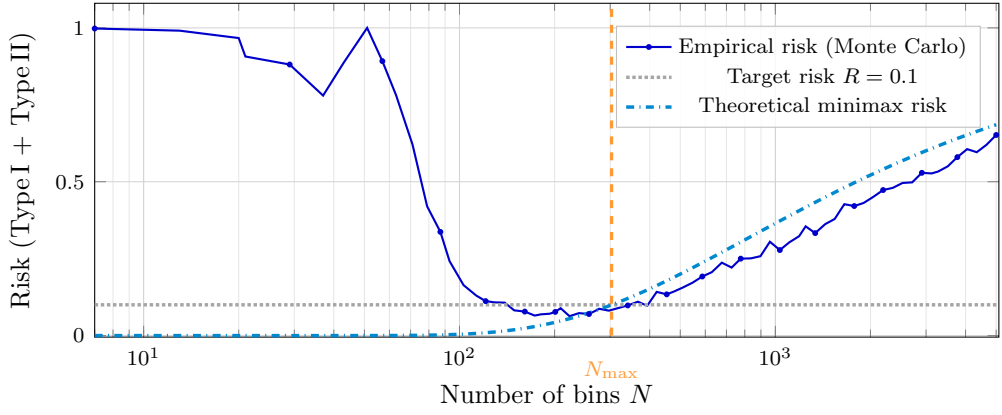
\begin{figure}[t]
    \centering
    \begin{tikzpicture}
    \begin{semilogxaxis}[
        width=0.86\linewidth, height=6cm,
        xmin=7, xmax=5100, ymin=-0.005, ymax=1.08,
        xlabel={Number of bins $N$},
        ylabel={Risk (Type\,I $+$ Type\,II)},
        xlabel style={font=\small}, ylabel style={font=\small},
        tick label style={font=\scriptsize},
        legend style={at={(0.98,0.78)}, anchor=east, font=\scriptsize,
            fill=white, fill opacity=0.85, text opacity=1, draw=gray!50},
        grid=both,
        minor grid style={line width=0.1pt, gray!20},
        major grid style={line width=0.2pt, gray!35},
        clip=false,
    ]
    \addplot [blue!80!black, thick, mark=*, mark size=0.7pt, mark repeat=4]
        table[col sep=comma, x=N, y=risk_emp]{risk_vs_N_data.csv};
    \addlegendentry{Empirical risk (Monte Carlo)}
    \addplot [gray!70, densely dotted, line width=1.3pt, domain=7:5100, samples=2] {0.1};
    \addlegendentry{Target risk $R=0.1$}
    \addplot [cyan!70!blue, dashdotted, line width=1.3pt]
        table[col sep=comma, x=N, y=risk_epsinf]{risk_vs_N_data.csv};
    \addlegendentry{Theoretical minimax risk}
    \addplot [orange!80, dashed, line width=1.2pt]
        coordinates {(303,0) (303,1.08)};
    \node [orange!80, below, font=\scriptsize] at (axis cs:303,-0.05) {$N_{\max}$};
    \end{semilogxaxis}
    \end{tikzpicture}
    \caption{Calibration-testing risk as a function of the number of bins $N$, for $n=5000$ test samples from the oscillatory density \eqref{eq:oscillatory-density} with $k=50$ and $a=0.2$, averaged over $1000$ Monte Carlo repetitions. Discretization bias dominates for small $N$, and statistical noise dominates for large $N$. The meeting point of the target risk $R=0.1$ with the theoretical minimax risk yields $N_{\max}=303$.}
    \label{fig:oscillatory-calibration}
\end{figure}

For example, \citet{kipnis2026calibrating} considers
\[
    n=5000,\qquad
    k=50,\qquad
    a=0.2,
\]
for which
\[
    \epsilon_\infty=\frac{2a}{\pi}\approx 0.127.
\]
For \(p=1\) and target minimax risk \(R=0.1\), the sharp-constant formula gives
\[
    N_{\max}=303.
\]
The resulting risk curve is illustrated in Figure~\ref{fig:oscillatory-calibration}.
With too few bins, bin averaging hides the
oscillatory miscalibration. Around the intermediate resolution, the oscillation
is resolved and the test has power. With too many bins, sampling noise dominates
and power is lost again.

This example illustrates why constants matter. A rate-level statement says that
testing is possible at the right asymptotic scale. The sharp-constant formula
says how fine the binning can be while still achieving a desired risk level.

\section{A broader lesson}

Uniformity testing is often considered a solved problem because the sample-complexity rate is known. Calibration testing is sometimes treated as an engineering problem because binning appears to be a practical detail. Both views miss the same point: constants matter when theory is used to choose a procedure.

Sharp constants do not merely refine asymptotic rates. They identify the effective Gaussian experiment hidden inside a high-dimensional testing problem. They explain why different statistics are optimal in different regimes. And they turn qualitative detectability statements into quantitative design rules.

From this perspective, uniformity testing is not only a classical benchmark problem. It is a laboratory for understanding how to design reliable tests in modern data-analysis pipelines.

Calibration binning is one such pipeline. The number of bins determines the resolution at which miscalibration can be seen, but also the statistical power with which it can be detected. Sharp constants provide a principled way to navigate this trade-off.

\section{Takeaway}

\[
    \textit{Rates tell us when testing is possible; constants tell us how well it works.}
\]

In distribution testing, this distinction matters because many natural procedures share the same rate but differ in risk. In calibration testing, it matters because binning choices determine whether a finite validation set can actually detect the miscalibration one hopes to measure.

Sharp constants therefore provide a bridge from asymptotic theory to statistical practice. They turn large-alphabet uniformity testing from a purely theoretical problem into a guide for choosing the resolution of calibration diagnostics.

\bibliographystyle{plainnat}
\bibliography{multinomials}

\begin{thebibliography}{5}
\providecommand{\natexlab}[1]{#1}
\providecommand{\url}[1]{\texttt{#1}}
\expandafter\ifx\csname urlstyle\endcsname\relax
  \providecommand{\doi}[1]{doi: #1}\else
  \providecommand{\doi}{doi: \begingroup \urlstyle{rm}\Url}\fi

\bibitem[Gupta and Price(2022)]{gupta2022sharp}
Shivam Gupta and Eric Price.
\newblock Sharp constants in uniformity testing via the huber statistic.
\newblock In \emph{Conference on Learning Theory}, pages 3113--3192. PMLR,
  2022.

\bibitem[Kipnis(2026{\natexlab{a}})]{kipnis2026calibrating}
Alon Kipnis.
\newblock Calibrating the calibration tester: Optimal binning and minimax
  calibration testing for continuous predictive models.
\newblock In \emph{Towards Trustworthy Predictions: Theory and Applications of
  Calibration for Modern AI}, 2026{\natexlab{a}}.
\newblock URL \url{https://openreview.net/forum?id=dy7XNC3W0g}.

\bibitem[Kipnis(2026{\natexlab{b}})]{kipnis2026minimax}
Alon Kipnis.
\newblock The minimax risk in testing uniformity over large alphabets under
  missing-ball alternatives.
\newblock \emph{IEEE Transactions on Information Theory}, 72\penalty0
  (3):\penalty0 1831--1849, 2026{\natexlab{b}}.
\newblock \doi{10.1109/TIT.2025.3646804}.

\bibitem[Nussbaum(1999)]{nussbaum1999minimax}
Michael Nussbaum.
\newblock Minimax risk: Pinsker bound.
\newblock \emph{Encyclopedia of Statistical Sciences}, 3:\penalty0 451--460,
  1999.

\bibitem[Paninski(2008)]{paninski2008coincidence}
Liam Paninski.
\newblock A coincidence-based test for uniformity given very sparsely sampled
  discrete data.
\newblock \emph{IEEE Transactions on Information Theory}, 54\penalty0
  (10):\penalty0 4750--4755, 2008.

\end{thebibliography}

\end{document}